\documentclass{ws-mpla}

\begin{document}

\markboth{Jihn E. Kim} {Self-tuning of cosmological constant and
exit from inflation}

%
%

\title{
Self-tuning of cosmological constant and exit from 
inflation
\footnote{Talk presented at {\it 2003 Int. Symposium on 
Cosmology and Particle Physics}(CosPA-03), National 
Taiwan Univ., Taiwan, Nov. 13--15, 2003.}
}


\author{\footnotesize Jihn E. Kim}

\address{School of Physics, Seoul National University,
Seoul 151-747, Korea
\\
jekim@phyp.snu.ac.kr}

\maketitle


\begin{abstract}
I review the recent 5D self-tuning solutions of the cosmological
constant problem, and try to unify two cosmological constant
problems within the framework of the self-tuning solutions. One
problem, the large cosmological constant needed for inflation, is
interpreted by starting with the parameters allowing only the dS
vacuum, and the vanishing cosmological constant at a true vacuum
is realized by changing parameters by exit from inflation at the
brane such that the self-tuning solution is allowed.

\end{abstract}

\ccode{PACS Nos.: 98.80.Es, 98.80.C, 12.25.Mj}

\section{Introduction}

In particle physics, there are three vacuum energies to deal with:
(i) the GUT scale vacuum energy for inflation, (ii) the vanishing
cosmological constant at the minimum which is theoretically
favored, and (iii) the observed tiny vacuum energy $(0.003\ {\rm
eV})^4$ at $z=10-100$. The tiny vacuum energy discovered from Type
1A supernova data and confirmed by the WMAP data makes the
cosmological constant problem more difficult. Here, we do not
discuss this tiny vacuum energy. Simply, we assume that somehow
this tiny vacuum energy is resolved by a (pseudo)scalar field
\cite{quint} whose misalignment shifts the vacuum energy a little
bit from a true minimum of the vanishing vacuum energy with
$\omega=p/\rho\le 0.8$. The cosmological constant($\Lambda=8\pi
G_NV_0$) is a term in Einstein's equation
\begin{equation}\label{Einstein}
R_{\mu\nu}-\frac12 Rg_{\mu\nu}-8\pi G_N V_0 g_{\mu\nu}=8\pi G_N
T_{\mu\nu}.
\end{equation}
The LHS of (\ref{Einstein}) without $\Lambda$ vanishes when the
spacetime is flat, $g_{\mu\nu}=\eta_{\mu\nu}$. Thus, if matter is
present, the RHS of (\ref{Einstein}) is nonvanishing and hence
spacetime is not flat, i.e. $g_{\mu\nu}\ne\eta_{\mu\nu}$.
Therefore, under a reasonable assumptions, Eq. (\ref{Einstein})
must lead to an evolving universe. In 1910's, the universe looked
as if it were a static one. So, Einstein introduced a compensating
term, $\Lambda g_{\mu\nu}$, on the LHS of (\ref{Einstein}) to make
the solution static. This is the birth of the cosmological
constant. In this talk, I will neglect matter(i.e. quantum
excitations), and hence we are looking for the vacuum solution.
Without matter but with a nonvanishing cosmological constant
$\Lambda$, it is not possible to have a static solution.

Probably an easy way to see the cosmological constant(c.c.)
problem is to start with an action where it is easy to find out
possible symmetries. The LHS of the Einstein equation with the
c.c. is obtained from the following action
\begin{equation}\label{action}
S=\int d^4x \sqrt{-g}\left(\frac{M^2}{2}R-V_0\right)
\end{equation}
where $g$ is the determinant of the metric tensor, $R$ is the
Ricci scalar, $M^2$ is the reduced Planck mass $M^2=1/8\pi G_N$,
and $V_0$ is a constant which is related to the cosmological
constant by $V_0=M^2\Lambda$.

The difficulty with the c.c. problem is that there is no symmetry
working for it to vanish. An obvious symmetry is the scale
invariance, but it is badly broken. At the electroweak scale, we
need a mass parameter of order 100~GeV, so the c.c. is expected to
be $10^{56}$ times larger than the current value. This
cosmological constant problem surfaced as a very severe one when
the spontaneous symmetry breaking of the electroweak theory was
extensively discussed~\cite{Veltman}.

The Einstein equation can be solved with an appropriate ansatz for
the metric. For example, a flat space ansatz is
$g_{\mu\nu}=\eta_{\mu\nu}$, and a de Sitter(dS) space ansatz is
\begin{equation}
ds^2=-dt^2+e^{2\sqrt{\Lambda} t}d{\bf x}^2.
\end{equation}
In 4D, the flat space is possible with the vanishing vacuum
energy. Thus, to have a flat space one has to fine-tune the
parameters in the action such that the vacuum energy is exactly
zero. To have a flat 4D space with a finite range of parameters in
the action, one must go beyond 4D.

Our 4D may come from higher dimensions. In this regard, the
Randall-Sundrum(RS) type models~\cite{rs2} are very interesting.
These are most easily studied in 5D. It will be sufficient if we,
the 4D observers, see the vanishing effective 4D c.c. even though
we start with nonzero 5D c.c. Indeed, the RS-II \cite{rs2} model
starts with a negative 5D c.c. $\Lambda_b$(bulk c.c.), $AdS_5$,
and a 3-brane(s) with brane tension $\Lambda_1$, but allow flat
4D, as shown by the line element
\begin{equation}
ds^2=\beta^2(y)(-dt^2+d{\bf x}^2) +dy^2
\end{equation}
where $\beta(y)$ is the warp factor. If there exists a reasonable
solution with the line element (4), then it describes a flat
space. It is usually assumed that matter is present at the brane
B1 located at $y=0$. Indeed, a flat 4D is possible with one
fine-tuning, $k_b=k_1$, with the exponentially suppressed warp
factor for large $y$; thus the 5th direction is not noticeable to
the 4D observer of B1 and there can results a consistent 4D. An
important thing to note is that a flat 4D is obtained, starting
with nonzero c.c. But still a fine-tuning between parameters is
needed.

\section{Self-tuning solutions}
\indent {\it Strong self-tuning solution}: Without fine-tuning
there exists a flat-space solution which is not continuously
connected to dS or AdS solutions. It was tried by Stanford groups
\cite{kachru}, but soon it has been shown that it has either the
singularity problem or reintroduces a fine-tuning \cite{FLLN}. The
solution they obtain has a singularity at say $y=y_c$. This is
because to cure it, or to satisfy the sum rule, one introduces
another brane at $y=y_c$. As soon as one introduces a brane, there
is one more parameter introduced there, namely the brane tension.
A flat space is possible only for a specific value of the new
brane tension, needing a fine-tuning. So there does not exists an
example for the strong self-tuning solution.

{\it Weak self-tuning solution}: This solution does not question
whether the flat solution is connected to dS or AdS solutions. Of
course, the flat self-tuning solution does not require a
fine-tuning(s). It was proposed in early 80's with antisymmetric
tensor field strength $H_{\mu\nu\rho\sigma}$. But because that was
done in 4D, the idea was not realistic. However, in 5D the weak
self-tuning solutions can be made realistic in RS-II models.

In RS-II type models, indeed there exist weak self-tuning
solutions \cite{kkl1,kl1}. One solution employs the antisymmetric
tensor field $A_{MNP}$ in 5D $AdS_5$ and one brane located at
$y=0$. The bulk c.c. is $\Lambda_b$ and the brane tension is
$\Lambda_1$. If one introduces the standard term $H^2$, there does
not exist a self-tuning solution. In fact there exists the no-go
theorem with a standard kinetic energy term \cite{nogo}. The
authors of Ref. [5] used $1/H^2$ term where
$H^2=H_{MNPQ}H^{MNPQ}$, and found a self-tuning solution. The
action is
\begin{equation}
S=\int d^4x dy\left\{\frac12 R_{(5)}-\frac{2\cdot 4!}{H^2}-
\Lambda_b-\Lambda_1\delta(y) \right\}
\end{equation}
where $R_{(5)}$ is the 5D Ricci scalar and we set the 5D
fundamental mass $M=1$. $\Lambda_b$ and $\Lambda_1$ define two
mass parameters: $k_b=\sqrt{-\Lambda_b/6},\ k_1=\Lambda_1/6$. The
self-tuning solution with the $Z_2$ symmetry condition is given by
\begin{equation}
\beta(y)=\left[\frac{a}{k_b}\cosh(4k_b|y|+c)\right]^{-1/4}
\end{equation}
where $a$ is an integration constant from the field equation of
$A_{MNP}$ and $c$ is another integration constant. Note that the
brane tension $\Lambda_1$ is not fine-tuned. But as we will
discuss, it must be bounded for the self-tuning solution to be
possible. $a$ is determined by the magnitude of the 4D Planck
mass, and $c$ is determined by the boundary condition at the brane
$-\beta(y)^\prime\Big|_{y=0^+}=\Lambda_1/6$ where $\beta$ is
normalized as $\beta(0)=1$. This self-tuning solution is connected
to dS and AdS solutions \cite{kkl1}; hence it is a weak
self-tuning solution. Since the solution is given in a closed
form, it can help studying some properties of weak self-tuning
solutions. The existence of the self-tuning solution in the RS-II
setup can be recognized by looking at the equation
\begin{equation}
|{\beta^\prime}|=\sqrt{{\bar k^2}+k_b^2\beta^2-a^2{\beta^{10}}},
\end{equation}
where $\bar k$ is the effective curvature of the 4D space, $\bar
k^2=+,0,-,$ corresponding to dS, flat and AdS, respectively. Note
that the positive power of $\beta$ in the $a^2$ term in Eq.~(7)
when one uses $1/H^2$. It would have a negative power $\beta^{-6}$
if one used $H^2$ instead. The flat solution needs
$\beta^\prime\rightarrow 0$ as $\beta\rightarrow 0$. Therefore,
the case $H^2$ does not satisfy this self-tuning solution
criterion. On the other hand, $1/H^2$ term satisfies this
condition. From this observation, one can find more self-tuning
solutions \cite{kl1}. Of particular interest among these is the
self-tuning solution from the log function $\log(H^2)$. Since the
solution with $1/H^2$ is given already, in the remainder of this
talk we focus on this solution given in (6). The boundary
condition at $y=0$ determines $c$,
\begin{equation}
\tanh (c)=\frac{\Lambda_1}{\sqrt{-6\Lambda_b}}.
\end{equation}

\section{Blowing up solutions}

The condition for the self-tuning solution, Eq. (8), is not always
satisfied. It is because the region of $\tanh$ is limited to
$[-1,+1]$. To see when it is not satisfied, let us note that the
boundary condition at $y=0^+$ requires
\begin{equation}
|\beta^\prime|_{0^+}=\sqrt{\bar
k^2+k_b^2-a^2}=\frac{\Lambda_1}{6}\ ,
\end{equation}
where $k_b^2=\frac{-\Lambda_b}{6}$. Therefore, the condition that
the flat solution is not allowed is
\begin{equation}
\frac{|\Lambda_1|}{6}>\sqrt{\frac{-\Lambda_b}{6}},
\end{equation}
or
\begin{equation}
\bar k^2>a^2\longrightarrow \overline{\Lambda}>
6a^2M^2\beta^{10}(0)
\end{equation}
where $\overline{\Lambda}$ is the effective curvature of the 4D
space. Thus, only the de Sitter space is allowed where the flat
space solution is forbidden \cite{kimjhep}. This situation is
shown in Fig. 1 (b) as the shaded regions. The light(dark) shaded
region is where $\Lambda_1$ is positive(negative). For comparison,
in Fig. 1 (a), we show the 4D case where the flat space is
possible only for $\Lambda_{\rm eff}=0$. This $\Lambda_{\rm
eff}=0$ is blowed up to a finite region, $-\sqrt{-6\Lambda_b}<
\Lambda_1<+\sqrt{-6\Lambda_b}$, in Fig. 1 (b). Note that the AdS
solution is also forbidden where the flat space solution is
forbidden.

\begin{figure}[th]
\centerline{\psfig{file=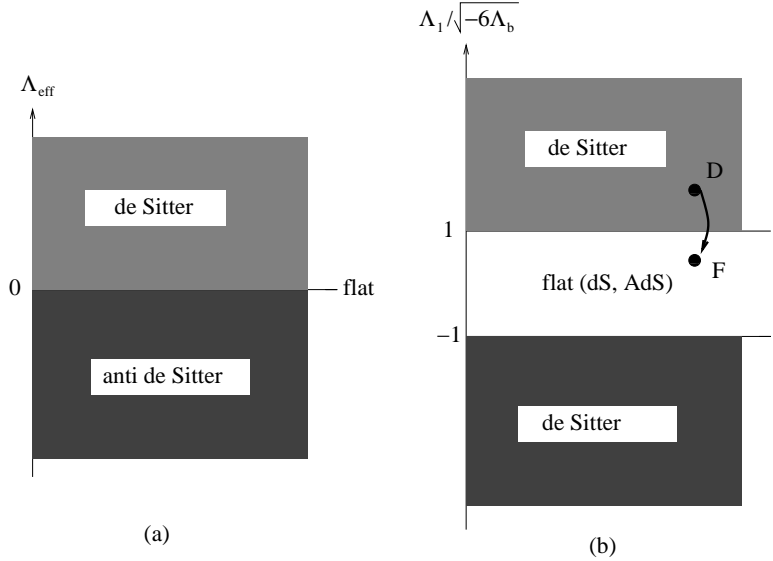,width=4.0in}} \vspace*{8pt}
\caption{A schematic illustration of the blow-up solution. The
universe at the dS phase(Point D) goes into the flat-allowed
region(Point F) by the parameter change at the brane.}
\end{figure}

The region of the dS-only region is distinguished by positive and
negative tensions. Because the negative tension leads to a phantom
with negative kinetic energy \cite{phantom}, we adopt the
parameters such that only a positive tension results.

This behavior of blowing up is a desirable property of weak
self-tuning solutions. For the $1/H^2$, it was possible to show
the blowing-up property as given above but it is difficult to see
it if one does not have a closed form solution.

\section{Inflation with self-tuning solution}

If only the de Sitter space is allowed for a finite range of
parameter space in the Lagrangian, it is tempting to use it for a
condition for inflation. By dynamics at the brane, we propose that
after a sufficient inflation the parameter settles to the region
where flat space solutions are possible. If this happens, say the
universe goes from D to F in Fig. 1 (b), the universe at F starts
from a de Sitter space solution. It seems to be possible because a
sudden change of the $\Lambda_1$ parameter would not change the
effective c.c. abruptly. But the F region also allows the flat
space, and the key question is how the flat space is chosen from a
dS solution after the sudden change of $\Lambda_1$. We hope to
find a reasonable history for this scenario, which unifies the
solutions of the large c.c.(for GUT phase inflation) and the
vanishing c.c.(by the self-tuning solution). This is the dream we
hoped to understand inflation on top of the vanishing
cosmological constant from the time of the birth of inflation
\cite{guth}.

For this purpose, let us adopt the hybrid inflation at the brane.
The key point of the hybrid inflation is to use multi fields among
which there are an inflaton field $\phi$ and a waterfall field
$\psi$ \cite{hybrid}. This setup is particularly relevant for our
scenario since at the brane $\Lambda_1$ can change
instantaneously. The coupled potential of the fields $\phi,\psi$
living on the brane is taken as
\begin{equation}
V= V_0+\frac12 m^2\phi^2 +\frac12 (-m_\psi^2+\lambda^\prime
\phi^2)\psi^2+\frac12 \lambda^\prime \psi^2\phi^2+\Lambda_1
\end{equation}
where $V_0=m_\psi^2\mu^2, m_\psi^2=\lambda\mu^2$, and we treat
$\mu^2$ and $m^2$ the large and small parameters, respectively.
For $\phi$ greater than the critical value
$\phi_c=\mu\sqrt{\lambda/\lambda^\prime}$, the waterfall field is
kept at origin $\psi=0$, and there results a slow roll inflation.
Requiring $m^2$ to be smaller than the Hubble parameter, $m^2\ll
H^2$, we obtain
\begin{equation}
\mu^2\gg \sqrt{\frac{12}{\lambda}}\ mM_{Pl}
\end{equation}
where $M_{Pl}=1.2\times 10^{19}$ GeV. The condition (13) also
guarantees a sufficient inflation. The condition for forbidding a
flat space solution is
\begin{equation}
\mu^4> \sqrt{\frac{96}{\xi \lambda^2}M_{Pl}^2
\sqrt{ak_b}|\Lambda_b|}
\end{equation}
where $\xi=O(1)$. For example, the eyeball numbers, $M= 8.4\times
10^{16}$ GeV, $\mu> 2.2\times 10^{15}$ GeV,
$V_0\sim\overline{\Lambda}> (1.86\times 10^{13}\ {\rm GeV})^4$,
satisfy these conditions. Thus, we can obtain reasonable numbers
from GUT related models. If $\phi$ crosses the critical value
$\phi_c$, the origin of $\psi$ becomes the vacuum for tachyonic
$\psi$, and the waterfall field $\psi$ runs into the true vacuum
immediately \cite{hybrid}. This parameter change occurs at the
brane, and the brane tension after the waterfall field settles at
the minimum is $\Lambda_1$. Namely, the universe settles at the
point F, with a nonzero brane tension
$\Lambda_1>0$, after the waterfall field finds the minimum. It has been
shown that the parameters of the above hybrid inflation can fall
in the region which GUT phase transitions allow \cite{kimjhep}.

\begin{figure}[th]
\centerline{\psfig{file=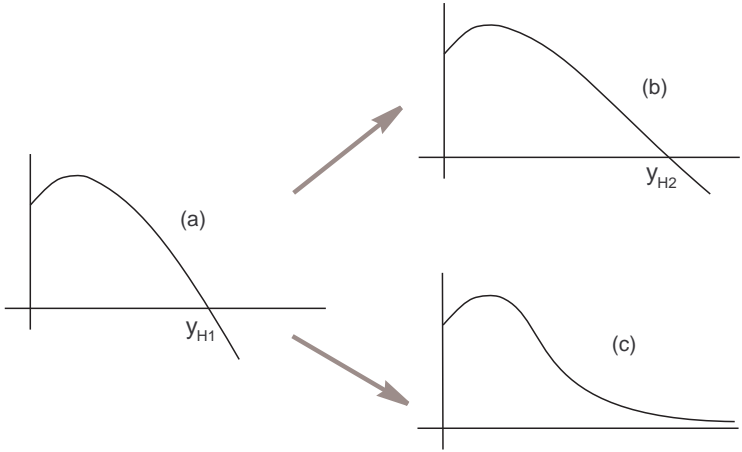,width=4.0in}} \vspace*{8pt}
\caption{A schematic illustration of the transition from a dS
space with horizon $H1$ (a) to another dS space with horizon $H2$
(b), or to a flat space (c).}
\end{figure}

Immediately after the hybrid inflation at the brane, 
our self-tuning solution is expected to be a dS one in the flat
allowed blowing-up region. Then, one can imagine that the dS
solution undergoes to a solution with a time-dependent c.c. But
the a closed form dS solution is not obtained. At the moment, the
best we can obtain is the existence of the time-dependent c.c. For
this, we studied a step function change of c.c. with a metric with
time-dependent $b(t)$ \cite{kl2}. As illustrated in Fig. 2, one
anticipates the solution with the curvature change, i.e. the
horizon distance changes from $y_{H1}$ (a) to $y_{H2}$ (b), 
or to $\infty$
(c). We found that in general there exist solutions for any 
$y_{H2}$.
Therefore, classical physics does not determine the path. Here,
Hawking's probabilistic interpretation \cite{hawking} 
is applicable. In our
setup, the initial condition after inflation is well defined, i.e.
Point F in Fig. 1 (b). Then, it is shown that the probability to
choose the flat space, i.e. Fig. 2 (c), is infinitely larger than
choosing any other space \cite{hawking}.

\section{Conclusion}

In conclusion,

\begin{romanlist}[(ii)]
\item We tried to unify two vacuum energy problems, one the
sufficient inflation and the other the vanishing cosmological
constant.
\item A weak self-tuning solution was used for this scenario for
the ultimately vanishing cosmological constant. It is possible for
some range of inflaton field parameters for inflation to happen,
$\Lambda_1>\sqrt{-\Lambda_b/6}$.

\item For inflation at the brane and the exit from inflation,
we showed the possibility in the hybrid inflation model.

\item The choice of the flat space after inflation is hoped to be
solved classically. If classical physics does not determine the
path,
quantum corrections must choose the ultimate late universe.

\end{romanlist}

\section*{Acknowledgments}

This work is supported in part by the KOSEF ABRL Grant No. 
R14-2003-012-01001-0, the BK21 program of Ministry of
Education, and Korea Research Foundation Grant No.
KRF-PBRG-2002-070-C00022.

\end{document}